# Experimental Demonstrations of Coherence de Broglie Wavelength for Scalable Superresolution with Near-perfect Fringe Visibility

Sangbae Kim[1,2] and Byoung S. Ham[1,3,*]
[1]Department of Electrical Engineering and Computer Science, Gwangju Institute of Science and Technology, 123 Chumdangwagi-ro, Buk-gu, Gwangju 61005, S. Korea
[2]QSIM+, 145 Anam-ro, Sungbuk-gu, Seoul 02841, S. Korea
[3]Department of Electrical and Computer Engineering, Oregon State University, Corvallis, OR 97331, USA
(Submitted on June 25, 2026; *bham@gist.ac.kr)

**Abstract:** Quantum sensing and metrology have been extensively investigated over the past several decades to surpass the classical shot-noise limit and approach the Heisenberg limit. The hallmark of N00N-state-based quantum sensing is superresolution, characterized by the interference fringe pattern $(1 + \cos N\varphi)$. However, practical implementations are severely constrained by the achievable photon number N, reduced fringe visibility, and vulnerability to photon loss. Recently, several coherence-based approaches without using N00N states have been explored as alternative routes to superresolution. Among them, the coherence de Broglie wavelength (CBW) approach is fully compatible with coherence optics. Here, we experimentally demonstrate CBW-based superresolution up to $N = 3$. In contrast to N00N-state-based schemes, the observed CBW fringes exhibit near-unity visibility that is essentially independent of N and remain robust against photon loss. Although CBW does not attain the Heisenberg-limited phase sensitivity, its phase sensitivity exhibits an N-fold enhancement over conventional classical interferometric approaches. These results suggest that CBW provides a practical and scalable platform for superresolution-based sensing and metrology.

## Introduction

Precision measurements have long been a central research topic across a wide range of scientific and technological fields, including spectroscopy, imaging, sensing, and metrology [1-6]. The performance of classical optical systems is fundamentally constrained by diffraction-limited resolution [7] and statistical measurement noise [8]. The diffraction limit restricts spatial and phase resolution to approximately half the wavelength of the probing light, while measurement sensitivity is bounded by the shot-noise limit (SNL) [7], which arises from Poissonian photon-number fluctuations [8]. Overcoming the diffraction limit is classically possible by using nonlinear optics-based atomic transitions [1] and measurements [9], even though SNL remains unchanged for phase sensitivity. Overcoming SNL is the primary motivation of quantum sensing and quantum metrology [5,6]. Quantum metrology exploits nonclassical resources, such as entanglement and squeezed states, to enhance measurement performance beyond classical bounds [10-12]. In particular, maximally path-entangled N00N states generate interference fringes with an effective wavelength of $\lambda/N$, commonly referred to as the photonic de Broglie wavelength (PBW) [13]. This unique feature enables superresolution and, in principle, phase supersensitivity approaching the Heisenberg limit through N-fold phase accumulation [11-18]. Despite these theoretical advantages, N00N-state-based approaches face significant practical challenges [16-18]. The generation and preservation of high-order entangled photon states remain experimentally demanding, while interference visibility deteriorates rapidly in the presence of photon loss, decoherence, or order mixing. As a result, the scalability and practical applicability of N00N-state-based quantum sensing remain severely limited [16-18].

An alternative route to superresolution has been proposed for non-quantum light, utilizing laser-attenuated single photons [19-24] and even continuous-wave (CW) light [22-24]. Under this non-quantum superresolution category, some methods are simply phase-control based [19,20], and others are phase accumulation-based [21,24]. In the latter, multi-pass [21,25] and cascaded interferometer [22-24] architectures are used for Nth power unitary transformation. Here, we report experimental demonstration of coherence de Broglie wavelength (CBW) generated from a cascaded Mach-Zehnder interferometer (MZI) [22-24]. Unlike PBW [11-18], CBW arises from a phase correlation [22-24] rather than photon correlation [14-18]. The CBW produces the same effective wavelength scaling analogous to PBW. Because CBW relies on classical coherence properties of optical fields, it is intrinsically robust against photon loss and order N selectivity, where the order N depends only on the number of coupled interferometers [24]. While CBW cannot achieve supersensitivity beyond the standard quantum limit, analytical studies have shown that it can still provide N-fold enhancement in phase sensitivity relative to the SNL [24]. This N-enhanced sensitivity gives a practical advantage over N00N-based quantum sensing limited by generation, measurement, and scalability, In addition, CBW provides several key advantages over PBW, including scalable order N dependence, near-unity fringe visibility independent of interference order, and strong invariance to photon loss. These features suggest that CBW may provide a practical pathway toward scalable superresolution sensing and metrology without the severe resource requirements associated with quantum-entangled states. A recent quantum mechanical analysis further clarified the physical origin of CBW and its relationship to higher-order interference phenomena in quantum optics [24].

**Results**

*Theory*

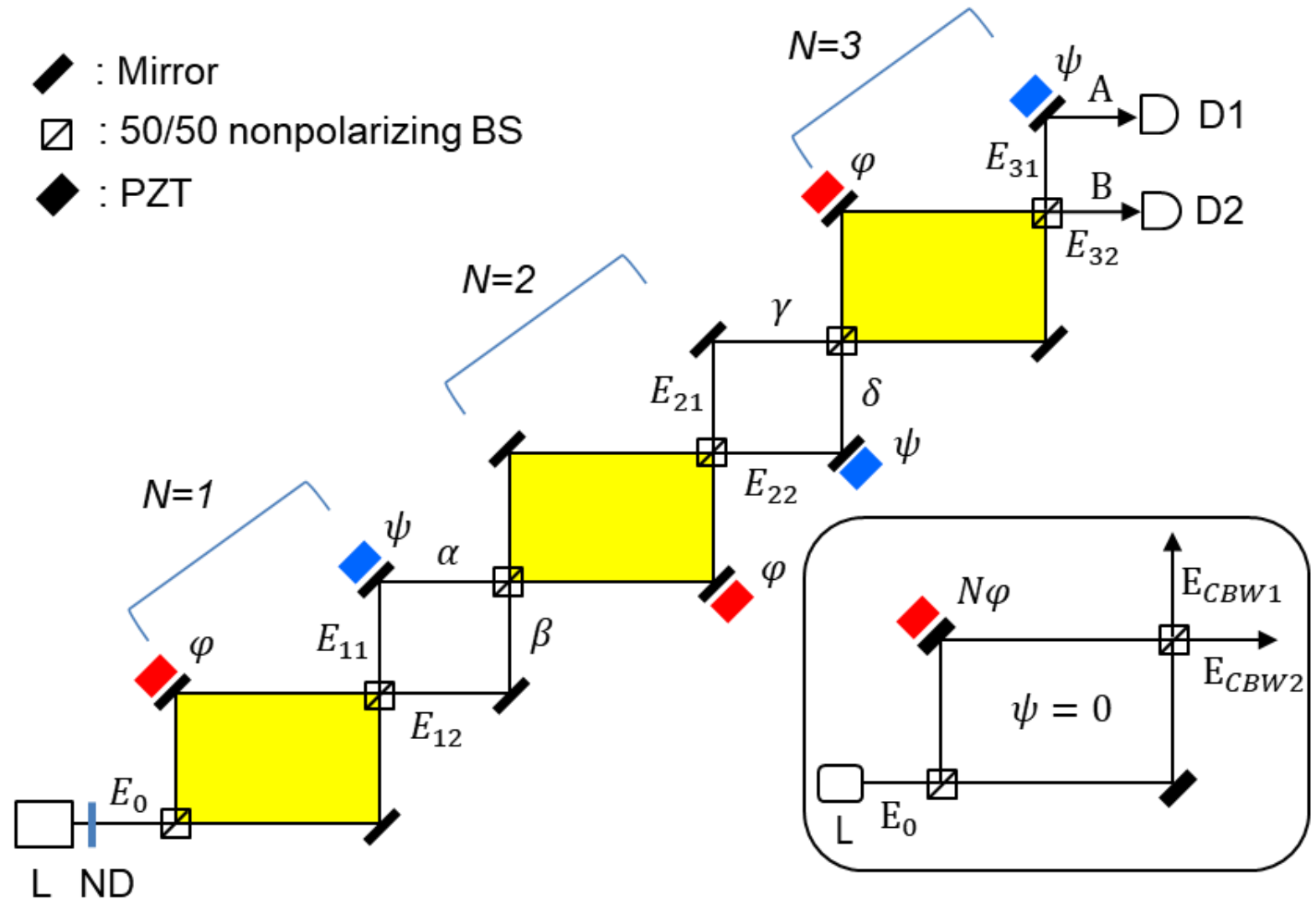

**Fig. 1. Schematic of Nth-order CBW.** L: Laser, ND: neutral density filters, D: single-photon detector, CCU: coincidence counting module. The input field $E_0$ is an attenuated continuous-wave laser at 100 μW. For CBW, the $\psi = 0$. Inset: Effective MZI for CBW sensing.

Figure 1 illustrates a schematic of the scalable CBW scheme implemented in a triply coupled MZI architecture, which enables the experimental achievement of PBW-like superresolution in the classical regime. The role of asymmetric coupling between adjacent MZIs indexed by $N = 1,2,3$ in the cascaded MZI structure is to match the same path basis of the MZIs, resulting in the *N*th power of the MZI unitary operator [24]. For this purpose, the dummy MZI ($\psi$) is essential to satisfy $D^{\dagger}\sigma_z D = \sigma_z$, where $\sigma_z$ is the Pauli operator, and D denotes the dummy MZI satisfying $[D, \sigma_z] = 0$. Here, the asymmetric coupling represents a π-phase shift between adjacent $\varphi$-MZIs [21]. The same asymmetry can also be achieved by controlling $\psi$ of the dummy MZIs for the identical $\varphi$-MZIs (see Section A of the Supplementary Materials).

The physics of CBW refers to the wave nature of a single photon in quantum mechanics; thus, CBW is intrinsically compatible with coherence optics. In contrast, the physics of PBW utilizes the particle nature of photons, where the corresponding superresolution is generated by photon-to-photon correlations of N00N states acting within the same single MZI. Although CBW differs fundamentally from PBW in its physical mechanism, these features can be understood in terms of the wave-particle duality in quantum mechanics. Unlike PBW-based quantum sensing, CBW superresolution is not affected by the photon loss in measurements. In addition, CBW exhibits near-perfect fringe visibility regardless of the order *N*. These features make CBW effective for implementing superresolution sensing with *N*-enhanced sensitivity, even though CBW remains within SNL [24].

The input field $E_0$ in Fig. 1 is provided by a commercially available laser (L). Neutral density filters (ND) are used to control the intensity of $E_0$ (see Methods). For the experimental demonstrations of CBW superresolutions, we first demonstrate scalable CBWs in a single-photon regime up to $N = 3$ (see Section B of the Supplementary Materials). Unlike PBWs, in which entangled photon pairs occupy both input ports of the first MZI, CBW uses only one input port, while the other port is open to vacuum noise. As analyzed quantum mechanically, CBW superresolution does not surpass SNL in the scaling of the random variable in measurements [24]. Secondly, we experimentally repeat the CBW superresolution using continuous-wave (CW) light, and the results are compared with those obtained from single-photon-based CBWs. Finally, analytic solutions and numerical calculations are performed to support the observed CBWs.

For the measurement of CBW superresolution in the single-photon regime, the coincidence-detection technique used in quantum sensing is adopted to isolate single photons from Poisson-distributed photon events (see Section B of the Supplementary Materials). As in N00N-state-based quantum sensing, the independent and identically distributed (*i.i.d.*) photons in measurements are ensured by a 50/50 nonpolarizing beam splitter (BS), whose output ports are randomly populated. Multiple photons arriving within the same detection time window of a single-photon detector may induce probabilistic errors unless photon-number-resolving detectors are used. Since the origin of this statistical error lies in the Poisson-distribution of photons, the intrinsic error in single-photon detection is typically only a few percent. Nevertheless, this statistical measurement error does not affect the observed CBW fringe visibility, because the MZI unitary transformation depends only on the relative phase $\varphi$, not on the input photon number or detected photon number. This property highlights the essence of coherence-optics-based CBW sensing and metrology.

To demonstrate the first-order ($N = 1$) CBW superresolution in Fig. 1, the output fields $E_{11}$ and $E_{12}$ are measured to obtain the first-order intensity correlation, resulting in normal MZI fringes as a reference for CBWs. These fringes arise from the unitary transformation of a single photon or the input field $E_0$, which is quantum mechanically represented by $e^{i\pi\sigma_x/2}$, where $\sigma_x$ is the Pauli operator [24]. It is known that each MZI behaves as a qubit in the CBW architecture, where the Pauli matrices generate the unitary operations acting on the path basis [24]:

$$U_{MZI}(\varphi) = BP(\varphi)B = e^{i\varphi/2}\begin{bmatrix} \cos\left(\frac{\varphi}{2}\right) & isin\left(\frac{\varphi}{2}\right) \\ isin\left(\frac{\varphi}{2}\right) & \cos\left(\frac{\varphi}{2}\right) \end{bmatrix}, \quad (1)$$

where $B = e^{i\pi\sigma_x/4}$ and $P(\varphi) = e^{i\varphi\sigma_z/2}$. In Eq. (1), the first-order CBW is expressed as $\lambda_{CBW}^{(1)} = \lambda_0$, where $\lambda_0$ is the wavelength of the input field. This is the general feature of classical sensors based on coherence optics.

For the second-order ($N = 2$) CBW superresolution, expressed as $\lambda_{CBW}^{(2)} = \lambda_0/2$, the output fields $E_{21}$ and $E_{22}$ are measured for the same first-order intensity correlation. However, in this case, the unit MZI transformation is doubled in the cascaded structure, resulting in a phase-doubling effect [24]:

$$U_{MZI}^{(2)}(\varphi) = \left(U_{MZI}(\varphi)\right)^{\otimes 2} = e^{i\varphi}\begin{bmatrix} \cos\varphi & isin\varphi \\ isin\varphi & \cos\varphi \end{bmatrix}. \quad (2)$$

In Eq. (2), the measured output fringes of $E_{21}$ and $E_{22}$ are doubled compared with those of $E_{11}$ and $E_{12}$ in Eq. (1). This origin of Eq. (2) is basis realignment between consecutive $\varphi$-MZIs by the $\psi$-MZIs [24]; otherwise, it results in an identity relation between the input and output [23,26]. In more detail, the same consecutive $\varphi$-MZIs are coupled through a π-phase shifted $\psi$-MZI. For the out-of-phase $\varphi$-MZIs in Fig. 1, a zero-phase shifted $\psi$-MZI is needed.

For the third-order ($N = 3$) CBW superresolution, the output fields $E_{31}$ and $E_{32}$ correspond to the tripled MZI unitary transformations, resulting in a tripled phase accumulation:

$$U_{MZI}^{(3)}(\varphi) = \left(U_{MZI}(\varphi)\right)^{\otimes 3} = e^{i3\varphi/2}\begin{bmatrix} \cos\left(\frac{3\varphi}{2}\right) & isin\left(\frac{3\varphi}{2}\right) \\ isin\left(\frac{3\varphi}{2}\right) & \cos\left(\frac{3\varphi}{2}\right) \end{bmatrix}. \quad (3)$$

Equation (3) shows the CBW superresolution for $N = 3$, resulting in $\lambda_{CBW}^{(3)} = \lambda_0/3$. Unlike Nth order intensity-correlation measurements in N00N-based superresolution for N photons [11-18], the measurement of the *N*th-order CBW superresolutions is for the first-order intensity correlation through N $\varphi$-MZIs, exhibiting N-independent near-unity fringe visibility (see Fig. 2).

To understand the uniqueness of CBW superresolution, as analyzed in Eqs. (1)-(3), the normal mode of a coupled classical system - such as a standard spring-mass chain - can be considered for comparison. In this case, the exact solution is $\omega_p = 2\sqrt{\frac{k}{m}} sin\left(\frac{p\pi}{2(N+1)}\right)$, where $m$ and $k$ denote the mass and spring constant, respectively, and $p = 1,2, \dots, N$. However, the resulting classical normal mode $\omega_p$ is not linear in N, although it may appear approximately linear for low orders. In fact, no classical counterpart of CBW superresolution has been identified.

*Experiments*

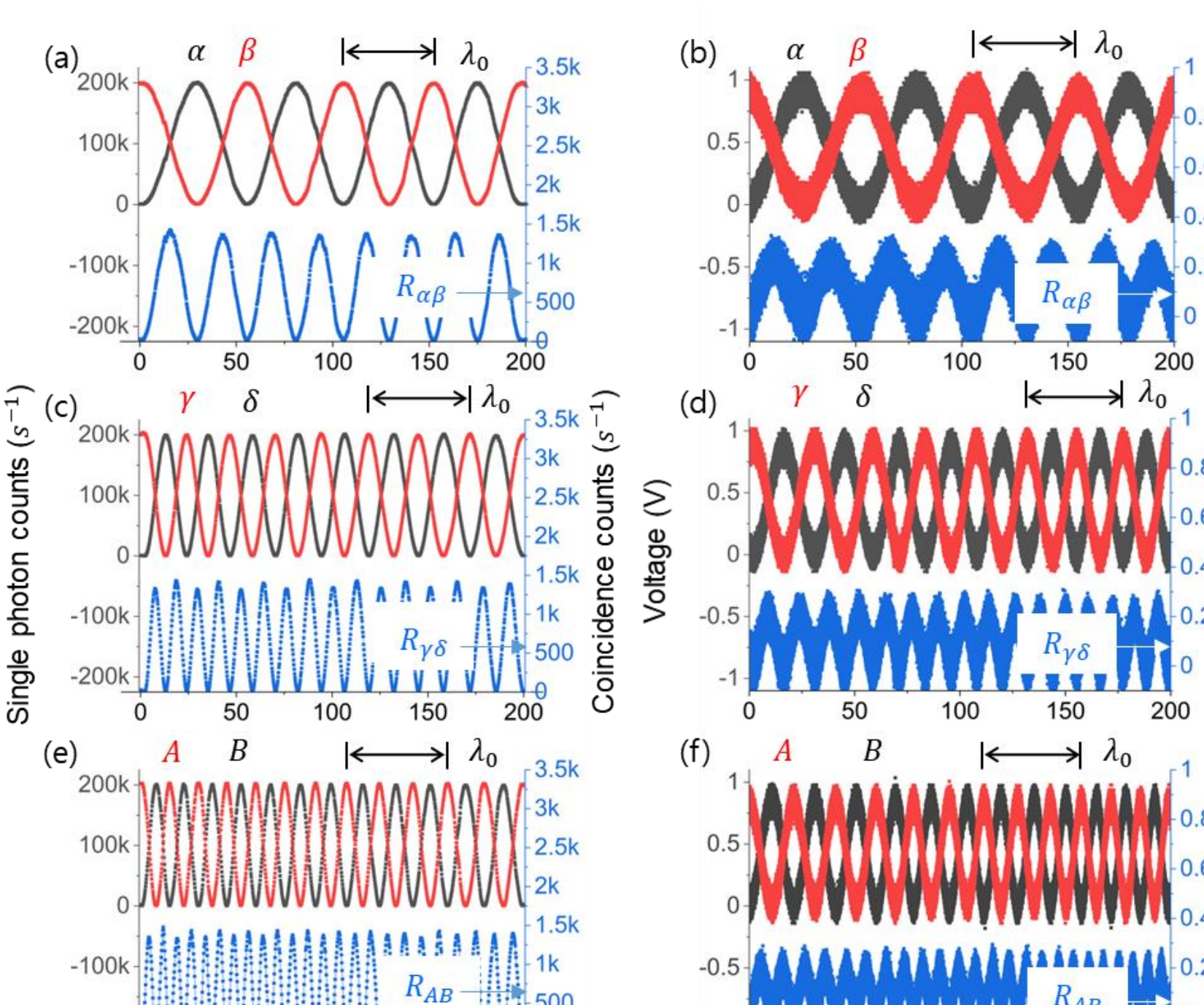

**Fig. 2. Experimental results of CBW in Fig. 1.** (a),(c),(e) single photon cases (see Methods). (b),(d),(f) CW cases at 100 μW CW power of 532 nm laser. The right vertical axis of each panel is for the coincidence count rate for $R_{ij}$, where *i* and *j* are the corresponding output pairs. The horizontal-axes are for the synchronized φs, where the phase of φs is controlled by PZT via a PZT controller (see Methods). From the top to the bottom rows, n=1, n=2, and n=3, respectively.

Figure 2 shows the experimental results of the CBW superresolution for $N = 1,2,3$ obtained from the asymmetrically coupled MZI architecture in Fig. 1 under the $\psi = 0$ condition of the dummy MZI. The left column of Fig. 2 corresponds to the single-photon regime, where different orders of CBWs are measured separately under a common scanning condition of synchronized piezoelectric transducers (PZTs), which is equivalent to the phase $\varphi$ in Eqs. (1)-(3) (see Methods). The right column of Fig. 2 corresponds to the CW cases with an input power of 0.1 mW for $E_0$, using the same PZT scanning condition as the single-photon case. As observed, both N-dependent fringes satisfy the superresolution relation $(1 \pm cos(N\varphi))/2$ with near-perfect fringe visibility (see details in Table 1). The fringe period $\lambda_0$ in the top row of both cases in Fig. 2 represents the wavelength of the input field $E_0$ as a reference. The coincidence measurement $R_{\alpha\beta}$ in the left column is directly read out from the coincidence counting unit (CCU; Altera DE2) for the second-order intensity correlation of the output signals α and β, detected by corresponding single-photon counting modules (Excelitas AQRH-SPCM-15). The corresponding $R_{\alpha\beta}$ in Fig. 2(b) for the CW case is obtained by calculating the product of the individual CW-output fields α and β. This calculation was performed in Excel due to the lack of dedicated hardware.

Figures 2(c) and 2(d) show the experimental results of the second-order ($N = 2$) CBW superresolution, where the fringe period is shortened to one half of that in Figs. 2(a) and 2(b). Thus, the second-order CBW superresolution is confirmed for $\lambda_{CBW}^{(2)} = \lambda_0/2$, with near-perfect fringe visibility (see Table 1). Likewise, Figs. 2(e) and 2(f) present the experimental results of the third-order ($N = 3$) CBW superresolution, satisfying $\lambda_{CBW}^{(3)} = \lambda_0/3$. Similarly, both the single-photon and CW CBWs exhibit near-perfect fringe visibility (see Table 1). Due to the wave nature of CBW, the photon-loss vulnerability, which is critical in PBW-based quantum sensing, is no longer relevant (not shown). Moreover, Fig. 2 demonstrates that a particular order of CBW is automatically selected by setting the number of φ-MZIs, owing to the asymmetrically coupled-MZI physics with SU(2) symmetry [24]. In contrast, N00N-state-based PBWs are inherently difficult to isolate from hyper-Poisson-distributed N00N states, resulting in inevitably imperfect fringe visibility [17,18]. Finally, the observed CBW superresolution in Fig. 2 demonstrates the quantized phase basis of CBW at $\varphi_N = m\pi/N$, where m=0,1,2,…,N (discussed below) [23]. This CBW phase quantization is mathematically equivalent to the Pegg-Barnett formalism of the phase operator [27]. Compared with PBWs, these features offer a significant advantage for a CBW-based superresolution sensing platform.

**Table 1. Measured CBW visibilities for Fig. 2.** The i (j) stands for α (β), γ (δ), and A (B) in each order in Fig. 2.

| Case \ Order | | $N = 1$ | $N = 2$ | $N = 3$ |
|---|---|---|---|---|
| Single photon | Red | $98.36 \pm 0.12$ | $99.51 \pm 0.07$ | $98.47 \pm 0.14$ |
| | Blue | $99.40 \pm 0.12$ | $99.56 \pm 0.01$ | $99.25 \pm 0.04$ |
| CW | Red | $100 \pm 0.0$ | $98.14 \pm 0.95$ | $100 \pm 0.0$ |
| | Blue | $98.88 \pm 0.52$ | $100 \pm 0.0$ | $97.40 \pm 0.83$ |

*Coherence solution of CBW*

From Fig. 1, the following analytical solutions are derived from MZI matrix representations:

$$\begin{bmatrix}E_A\\E_B\end{bmatrix} = [\varphi^+][M^+]\begin{bmatrix}E_\gamma\\E_\delta\end{bmatrix}, \tag{4}$$

where $\begin{bmatrix}E_\gamma\\E_\delta\end{bmatrix} = [\varphi^-][M^-]\begin{bmatrix}E_\alpha\\E_\beta\end{bmatrix}$, $\begin{bmatrix}E_\alpha\\E_\beta\end{bmatrix} = [\varphi^+][M^+]\begin{bmatrix}E_0\\1\end{bmatrix}$, $[M^+] = \frac{1}{2}\begin{bmatrix}(1-e^{i\varphi}) & i(1+e^{i\varphi})\\ i(1+e^{i\varphi}) & -(1-e^{i\varphi})\end{bmatrix}$, $[M^-] = \frac{1}{2}\begin{bmatrix}-(1-e^{i\varphi}) & i(1+e^{i\varphi})\\ i(1+e^{i\varphi}) & (1-e^{i\varphi})\end{bmatrix}$, $[\varphi^+] = \begin{bmatrix}1 & 0\\ 0 & e^{i\varphi}\end{bmatrix}$, and $[\varphi^-] = \begin{bmatrix}e^{i\varphi} & 0\\ 0 & 1\end{bmatrix}$. For the coherence optics-based analytic solutions, $\psi = 0$ is set for all dummy MZIs.

From Eq. (4), the following CBW relations can be derived:

$$\begin{bmatrix}E_\alpha\\E_\beta\end{bmatrix} = \frac{1}{2}\begin{bmatrix}(1-e^{i\varphi}) & i(1+e^{i\varphi})\\ i(1+e^{i\varphi}) & -(1-e^{i\varphi})\end{bmatrix}\begin{bmatrix}E_0\\0\end{bmatrix}, \tag{5}$$

$$\begin{bmatrix}E_\gamma\\E_\delta\end{bmatrix} = \frac{1}{4}\begin{bmatrix}-(1-e^{i\varphi}) & i(1+e^{i\varphi})\\ i(1+e^{i\varphi}) & (1-e^{i\varphi})\end{bmatrix}\begin{bmatrix}(1-e^{i\varphi}) & i(1+e^{i\varphi})\\ i(1+e^{i\varphi}) & -(1-e^{i\varphi})\end{bmatrix}\begin{bmatrix}E_0\\0\end{bmatrix}, \tag{6}$$

$$\begin{bmatrix}E_A\\E_B\end{bmatrix} = \frac{1}{8}\begin{bmatrix}(1-e^{i\varphi}) & i(1+e^{i\varphi})\\ i(1+e^{i\varphi}) & -(1-e^{i\varphi})\end{bmatrix}\begin{bmatrix}-(1-e^{i\varphi}) & i(1+e^{i\varphi})\\ i(1+e^{i\varphi}) & (1-e^{i\varphi})\end{bmatrix}\begin{bmatrix}(1-e^{i\varphi}) & i(1+e^{i\varphi})\\ i(1+e^{i\varphi}) & -(1-e^{i\varphi})\end{bmatrix}\begin{bmatrix}E_0\\0\end{bmatrix}. \tag{7}$$

Equation (5) yields $I_\alpha = I_0(1 - cos\varphi)/2$ and $I_\beta = I_0(1 + cos\varphi)/2$, where $I_j = E_j E_j^*$. These correspond to the typical MZI interference fringes, equivalent to the first-order CBW superresolution. Likewise, Eq. (6) yields $I_\gamma =$ $I_0(1 + \cos(2\varphi))/2$ and $I_\delta = I_0(1 - \cos(2\varphi))/2$, indicating the second-order CBW superresolution. Similarly, $I_A = I_0(1 - \cos(3\varphi))/2$ and $I_B = I_0(1 + \cos(3\varphi))/2$ are obtained from Eq. (7) for the third-order CBW superresolution. Thus, the general solution of the Nth-order CBW superresolution can be expressed as:

$$\begin{bmatrix} E_A \\ E_B \end{bmatrix}^{(N)} = (-1)^N \left(\frac{1}{2}\right) \begin{bmatrix} 1 + (-1)^N e^{iN\varphi} & -i(1 + (-1)^{N+1} e^{in\varphi}) \\ i(1 + (-1)^{N+1} e^{iN\varphi}) & 1 + (-1)^N e^{iN\varphi} \end{bmatrix} \begin{bmatrix} E_0 \\ 0 \end{bmatrix}. \quad (8)$$

The corresponding fringes of the Nth-order CBW superresolution can be directly obtained from Eq. (8):

$$I_A{}^{(N)} = \frac{1}{2}[1 + (-1)^N \cos(N\varphi)], \quad (9)$$

$$I_B{}^{(N)} = \frac{1}{2}[1 - (-1)^N \cos(N\varphi)]. \quad (10)$$

Equations (9) and (10) agree with the observed data in Fig. 2, satisfying the effective wavelength relation $\lambda_{A;B} = \frac{\lambda_0}{N}$. As a result, the Nth-order coincidence measurement becomes $R_{AB}{}^{(N)} = I_A{}^{(N)} \cdot I_A{}^{(N)} = \frac{1}{4} sin^2(N\varphi)$, resulting in doubled fringes (see Fig. 3).

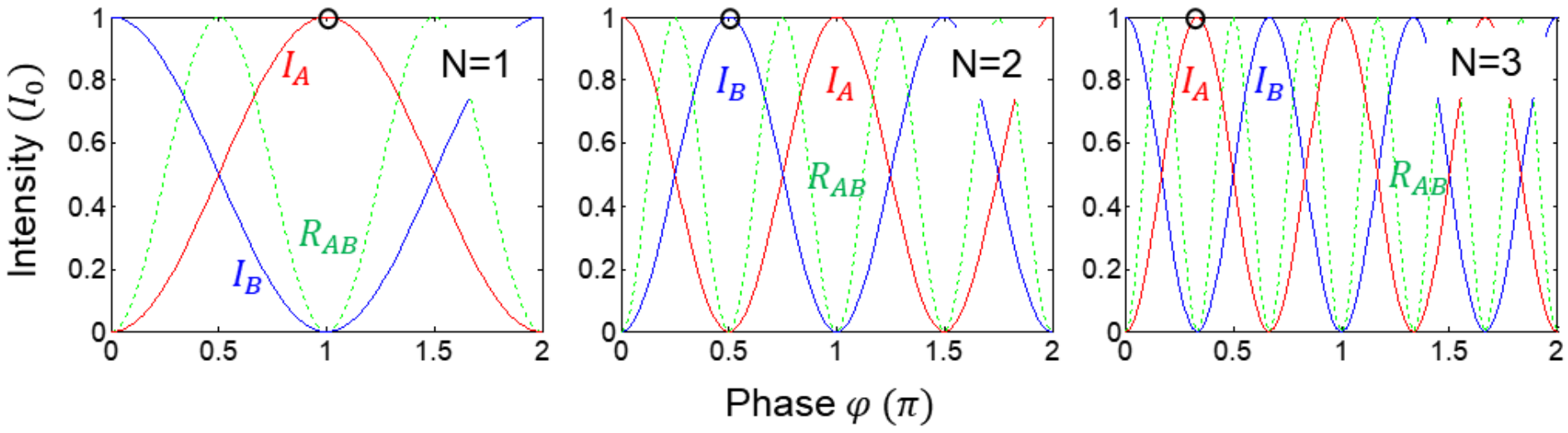

**Fig. 3. Numerical calculations of CBWs for Fig. 2.** The N represents the number of the basic building block defined in Fig. 1. The open circle indicates the phase basis quantization of the N-dependent CBWs. The intensity product $R_{AB}$ is normalized.

Figure 3 shows the corresponding numerical calculations of CBWs at $\lambda_{CBW} = \frac{\lambda_0}{N}$, obtained using Eq. (8), to support the experimental data in Fig. 2. The $(-1)^N$ factor in Eq. (8) appears in the ordered CBWs with a new phase basis $\varphi_N = m\pi/N$, where the phase-basis step is $\Delta\varphi_N = \pi/N$, as marked by open circles. This leads to an N-fold increase in resolution according to the Rayleigh criterion. The factor $(-1)^N$ in Eq. (8) can be interpreted as the eigenvalue of the parity operator $\hat{P} = (-1)^N$ (discussed elsewhere) [28,29]. Its eigenvalues, $\pm 1$, correspond to two quantized phase states $0$ and $\pi/N$, forming a phase basis of the *N*-dependent CBW [23]. The open circles in Fig. 3 indicate these phase-basis eigenstates, whose parity-dependent sign determines the constructive output port and results in the N-fold increase in phase resolution observed in Fig. 2. The intensity product $R_{AB}{}^{(N)}$ confirms the analytic solution $sin^2(N\varphi)$, resulting in twice the resolution given by $\lambda_{CBW}$ (see the green dotted curves). Although CBW enables a new superresolution sensing platform with near-perfect visibility and photon-loss invariance, its phase sensitivity remains within the SNL of classical physics, even though the corresponding Fisher information enhances by $N^2$ [24]. Thus, the novelty of the observed CBW superresolution lies in two aspects: first, its compatibility with classical sensing platforms due to its wave nature; and second, its practical advantages over N00N-state-based quantum sensing, which suffers from limited-order scalability [14-18], practically inseparable order N [18], and photon-loss vulnerability [10-18].

**Discussion**

For a practical sensitivity comparison between CBW using M-coupled MZIs and PBW using an $N_{N00N}$-photon N00N state, the ideal N00N-state sensitivity must satisfy $\Delta\phi_{PBW} < \Delta\phi_{CBW}$. Here, the CBW sensitivity is given by $\Delta\phi_{CBW} \sim \frac{1}{M\sqrt{N_{ph}}}$ [24], whereas the ideal PBW sensitivity is $\Delta\phi_{PBW} \sim \frac{1}{N_{N00N}}$. Thus, to overcome CBW, the N00N state must satisfy $\frac{1}{N_{N00N}} < \frac{1}{M\sqrt{N_{ph}}}$ [24]. For 0.1 mW input power, corresponding to $\sim 10^{14}$ photons/s, this condition requires $N_{N00N} > 10^7 M$. To date, the largest experimentally demonstrated genuine optical N00N state contains only $N_{N00N} = 5$ photons [30]. Although much larger photonic entangled states have been reported, they are GHZ or hyperentangled state [16] rather than path-entangled N00N states. Therefore, the requirement $N_{N00N} > 10^7 M$ is far beyond current quantum sensing capability. Even in the single-photon regime, where $N_{ph} \sim N_{N00N}$, as in single-photon array-detector LiDAR systems developed to overcome the electronic noise of CCD-based classical sensing platform [31], the condition becomes $N_{N00N} > M^2$. Therefore, CBW with only a few coupled MZIs can outperform practical N00N-state-based quantum sensing. For example, with *M*=3, PBW would require $N_{N00N} > 9$, already exceeding the demonstrated $N_{N00N} = 5$. This practical scalability constitutes the novelty of CBW-based sensing, which provides an M-fold sensitivity enhancement over the SNL while achieving PBW-like superresolution without requiring high-order entangled photon states.

**Conclusion**

In conclusion, we experimentally demonstrated scalable (N = 1,2,3) CBW superresolution in a triply coupled asymmetric MZI-chain system, where the observed superresolution resembles PBWs in quantum sensing. Unlike PBWs, whose superresolution arises from photon correlations within a single MZI and suffers from imperfect fringe visibility and photon-loss vulnerability, the observed CBW supreresolution exhibits order-independent, near-perfect fringe visibility due to the wave nature of quantum mechanics. Most importantly, the scalable CBW superresolution is inherently invariant to the photon loss owing to its coherence-based mechanism. The coherent observation of CBW superresolution in both the single-photon and CW regimes has not been previously reported. Thus, the observed CBW superresolution is fundamentally different from both classical and conventional quantum sensing platforms. For the single-photon-based CBW experiment, a commercially available 532 nm CW laser was optically attenuated to achieve a mean photon number of $\langle n \rangle \sim 0.04$, where the single-photon regime was verified using the coincidence-detection technique commonly employed in quantum sensing. Even with CW laser light, the same near-perfect fringe visibility was experimentally observed, confirming the CBW physics governed by the N-powered MZI unitary transformation. Analytical solutions were also derived to discuss CBW superresolution using a parity operator, leading to the CBW phase quantization. Under the wave nature of a single photon, the order-dependent phase quantization of CBW is analogous to the energy quantization associated with the particle nature of a single photon. Owing to its determinacy and scalability with order *N*, CBW has the potential to overcome the practical limitations of N00N-state-based quantum sensing and may enable CBW-based superresolution sensing platforms beyond $N > 100$. Such platforms could outperform classical counterparts, such as wavemeters and remote sensing systems, even in photon-loss regimes relevant to modern LiDAR technologies [32,33].

**Methods**

For the single photon-based CBW experiments, the input field $E_0$ was obtained by optically attenuating a 532 nm laser (Coherent, V-10) with an initial power of 100 μW using neutral density filters with an optical density (OD) of 13. As a result, the measured mean photon number was $\langle n \rangle \sim 0.04$ with respect to the dead time of the single photon detector (Excelitas AQRH-SPCM-15). This mean-photon number was confirmed using both a coincidence counting unit (CCU, Altera DE2) and a fast digital oscilloscope (Yokogawa DL9040; 500 MHz) via a pair of the single photon detectors (see Section B of the Supplementary Materials). For the single-photon CBW experiment, we therefore adopted a coincidence-detection method to suppress the majority of vacuum events. According to photon statistics, however, multiply bunched photons cannot be completely eliminated by this technique, resulting in a statistical error at a few percent in the single photon detection scheme. For the CW CBW experiment, a typical measurement method used in coherence optics was adopted, replacing the coincidence-detection module with avalanche photodiodes (Thorlabs APD 110A). For the PZT scan time of 200 s in Fig. 2, a total of 2.5 million data points were collected from the CCU. Because the PZT was scanned continuously, each data point represents an averaged accumulated over approximately 80 μs of the PZT scan time.

**Acknowledgment**

BSH gratefully acknowledges Prof. Ben Lee at Oregon State University for hosting the author during the sabbatical leave and for kindly providing office space in the department.

**Data availability**
All data generated or analyzed during this study are included in the published article.

**Funding**
BSH acknowledges that this work was supported by the IITP-ITRC grant (IITP 2026-RS-2021-II211810) funded by the Korean government (Ministry of Science and ICT). BSH also thanks the financial support from GIST Research Project grant funded by the GIST 2026.

**Author contribution**
B.S.H. conceived the idea, analyzed the data, and wrote the manuscript. S.K. conducted the experiments. Correspondence and request of materials should be addressed to BSH (email: bham@gist.ac.kr).

**Competing interests**
BSH is the founder of Qu-Lidar.

**Supplementary information** is available in the online version of the paper. Reprints and permissions information is available online at www.nature.com/reprints.